\documentclass[10pt,twocolumn]{article}

\usepackage[letterpaper,margin=0.75in]{geometry}
\usepackage{newtxtext,newtxmath}
\usepackage{graphicx}
\usepackage{booktabs}
\usepackage{amsmath}
\usepackage{authblk}

\usepackage[colorlinks,urlcolor=blue,linkcolor=blue,citecolor=blue]{hyperref}
\expandafter\def\expandafter\UrlBreaks\expandafter{\UrlBreaks\do\/\do\*\do\-\do\~\do\'\do\"\do\-}
\usepackage{comment}

\newcommand{\revised}[1]{#1}

\usepackage{multirow}
\usepackage{ragged2e}
\usepackage{tikz}
\usetikzlibrary{matrix}
\usepackage{pgfplots}
\pgfplotsset{compat=1.15}
\usepgfplotslibrary{groupplots}

\begin{document}


\title{Performance Analysis of Digital Processing-in-Memory through a Case Study on Convolutional-Neural-Network Acceleration}

\author{Orian Leitersdorf}
\author{Ronny Ronen}
\author{Shahar Kvatinsky}

\affil{Technion -- Israel Institute of Technology, Haifa, 3200003, Israel}

\date{}

\maketitle

\begin{abstract}\justifying\looseness-1Processing-in-Memory (PIM) architectures are evolving to minimize data movement by leveraging the same physical devices for both memory and logic functionalities. While analog PIM harnesses crossbar arrays for efficient approximate matrix-vector multiplication, digital PIM architectures facilitate massively-parallel bitwise operations for more general workloads. Recent works have extended digital PIM towards the full-precision acceleration of convolutional neural networks (CNNs), yet a comprehensive comparison with GPUs remains a gap in the literature that may illuminate the limitations of digital PIM. This paper aims to fill this void by conducting a thorough examination of CNN acceleration through an updated quantitative comparison with GPUs. Our approach begins with a theoretical investigation into various PIM architectures, shedding light on their performance characteristics and constraints. Subsequently, through a series of benchmarks spanning memory-bound vectored arithmetic to CNN acceleration, we provide insights into digital PIM performance that may guide the acceleration of applications in the future.
\end{abstract}

\section{Introduction} The \emph{memory wall} serves as a fundamental bottleneck to computing systems for \emph{memory-intensive} applications as the memory access occasionally becomes orders of magnitude more expensive than the computation itself~\cite{StatefulLogicReview}. Therefore, processing-in-memory (PIM) solutions aim to tackle the memory wall challenge by integrating processing capabilities directly into the memory architecture. That is, the traditional read/write interface is supplemented with \emph{logic} operations that perform vectored computation within the memory without transferring the data through the memory wall bottleneck. While early proposals for PIM integrated small processors within the memory, recent solutions perform logic by exploiting the physical properties of the memory devices themselves~\cite{StatefulLogicReview}. 

\begin{figure}[!t]
\centering 
\includegraphics[width=\linewidth, trim={0cm, 0.2cm, 0cm, 0cm}]{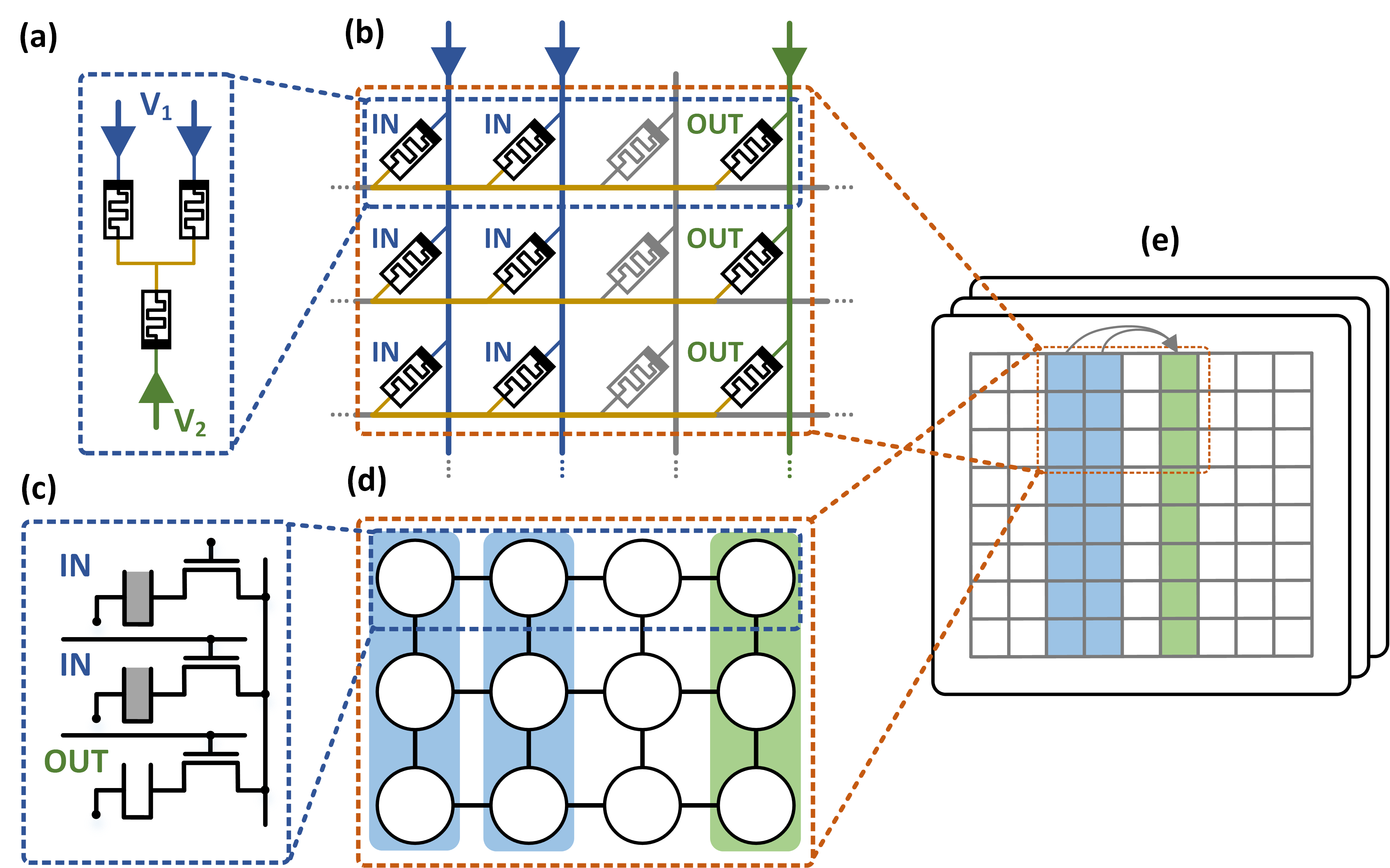}
\caption{Examples of digital PIM using (a, b) memristive~\cite{StatefulLogicReview} and (c, d) DRAM~\cite{SIMDRAM} memories. Both follow (e) an abstract model of bitwise column operations in $O(1)$ time. Figure adapted from~\cite{AritPIM}.}
\label{fig:PIM} 
\end{figure}

Numerous emerging PIM architectures exploit these physical properties towards \emph{digital bitwise} logic within memory arrays. For example, memristive stateful logic~\cite{StatefulLogicReview} utilizes the memristor, an emerging physical device with variable resistance, to perform logic in the resistive domain on binary values (e.g., low resistance is logical one and high resistance is logical zero). For memristors connected as seen in Figure~\ref{fig:PIM}(a), applying fixed voltages at the memristor terminals causes the resistance of the output memristor to become conditional on the resistances of the input memristors (e.g., the logical NOR of the inputs). As this circuit appears within every row of a crossbar array of memristors (see Figure~\ref{fig:PIM}(b)), we find that applying fixed voltages on the bitlines of the array can simultaneously induce a logic gate within every row of the crossbar array. Abstractly, we can consider a crossbar array as a binary matrix of memory, and stateful logic enables logic operations on columns of bits with $O(1)$ time (e.g., NOR of two columns into a third column), see Figure~\ref{fig:PIM}(e). Furthermore, this abstract model covers additional architectures such as in-DRAM computing~\cite{SIMDRAM}, where majority and negation gates may be performed in parallel across several columns. This bitwise parallelism can be exploited towards high-throughput arithmetic (e.g., addition, multiplication) for both fixed-point and floating-point numbers within the memory in a bit-serial element-parallel fashion: the logic gates that construct the arithmetic function are performed serially, yet in parallel across all rows of an array for parallel vectored execution. This can provide massive throughout and energy benefits over traditional hardware such as GPUs \emph{when the cache locality is low} (e.g., vectored arithmetic on vectors stored only in the main memory)~\cite{AritPIM}. 

\begin{table*}[t]
    \centering
    \caption{Summary of the evaluation parameters for GPU and PIM systems.}
    \begin{tabular}{|c|l|}
        \hline
        \textbf{Configuration} & \textbf{Parameters} \\
        \hline
        \hline
        \multirow{5}{*}{A6000 / A100 GPU} & \emph{Number of Cores:} 10752 / 6912\\
        & \emph{Memory Size:} 48 GB / 80 GB \\
        & \emph{Memory Bandwidth:} 768 GB/s / 1935 GB/s\\
        & \emph{Clock Frequency:} 1410 MHz / 1065 MHz \\
        & \emph{Max Power:} 300W / 300W \\
        \hline
        \multirow{5}{*}{Memristive / DRAM PIM} & \emph{Crossbar:} $1024 \times 1024$ / $65536 \times 1024$\\
        & \emph{Memory Size:} 48 GB \\
        & \emph{Gate Energy:} 6.4 fJ / 391 fJ \\
        & \emph{Clock Frequency:} 333 MHz / 0.5 MHz \\
        & \emph{Max Power:} 860 W / 80 W \\
        \hline
    \end{tabular}
    \label{tab:params}
\end{table*}

Recent works~\cite{FloatPIM, ReHy, 9181003, 9729451} have proposed utilizing digital PIM architectures towards convolutional neural network (CNN) acceleration. While emerging \emph{analog PIM} approaches~\cite{ProcIEEEAnalog} provide significant acceleration over GPUs \revised{due to their natural capability for matrix-vector multiplication}, they suffer from low accuracy due to noise in the analog domain and high costs for conversion between digital and analog domains. By supporting full-precision floating-point computation, digital PIM approaches have the potential to overcome these flaws for reliable CNN acceleration. FloatPIM~\cite{FloatPIM} was the first such work -- presenting vast improvement over GPU performance; several additional recent works~\cite{ReHy, 9181003, 9729451} have since built upon the FloatPIM architecture and evaluation. Yet, there is a need for an updated comprehensive evaluation of these works and a comparison to state-of-the-art hardware as:
\begin{enumerate} 
\item FloatPIM utilized routines that were either erroneous (e.g., the floating-point addition only supported unsigned numbers~\cite{ReHy, AritPIM}) or have been since improved (e.g., convolution~\cite{MatPIM}).
\item The GPU baseline considered in FloatPIM (and later utilized in the additional works) stores the weights in the CPU memory~\cite{FloatPIM}. In this paper, we demonstrate that storing the weights in the GPU memory significantly improves the GPU baseline.
\end{enumerate}
This paper aims to both provide an updated independent evaluation of CNN acceleration with digital PIM, and to highlight the underlying limitations of digital PIM architectures while giving further insight into new applications. The paper is structured as follows. The paper begins with the evaluation methodology of comparing memristive and DRAM PIM architectures to the NVIDIA A6000 and A100 GPUs. The paper then progresses through several benchmarks in the following sections, first considering routines utilized in CNN acceleration (high-throughput arithmetic, matrix multiplication and convolution) and then various large-scale CNN models for inference and training. We analyze the unique factors involved in each benchmark and develop several metrics that provide further insight into the performance of digital PIM architectures. Lastly, we discuss the implications of these results, revealing the key characteristics that may indicate the effectiveness of digital PIM acceleration for future applications.

\section{Methodology}
\vspace{2pt}

This section details the evaluation methodology that is utilized throughout this paper to compare digital PIM and GPU architectures. The PIM and GPU configurations utilized are summarized in Table~\ref{tab:params} \revised{and lead to four independent configurations that will be contrasted in this work: (1) memristive PIM, (2) DRAM PIM, (3) A6000 GPU, and (4) A100 GPU. The PIM architectures execute the entire processing with no additional memory or processing units (e.g., CPU/GPU).} 
The code repository\footnote{Available at \url{https://github.com/oleitersdorf/ConvPIM}.} includes (1) further details regarding the experiments that are necessary for their replication, and (2) additional results that comprise a sensitivity analysis across different GPUs, PIM configurations, and representation sizes. Overall, we find that those additional results strengthen the overall trends and insights discussed in this paper.

\subsection{GPU}

The GPU analysis provided throughout this paper is based on experimental and theoretical results from an NVIDIA A6000 GPU (workstation), supplemented with additional results provided in the code repository for the NVIDIA A100 GPU (datacenter). The A6000 (A100) GPU consists of 84 (108) independent Streaming Multiprocessor (SM) units, each can execute up to 128 (64) threads in parallel, for a total of 10752 (6912) cores overall. The experimental results are derived from the PyTorch\footnote{Available at \url{https://pytorch.org}.} library for general-purpose neural network acceleration, utilizing the built-in PyTorch profiler connected to the NVIDIA Nsight Systems profiler for GPU metrics (e.g., DRAM bandwidth, L1/L2 hit rate) and NVIDIA NVML\footnote{Available at \url{https://developer.nvidia.com/nvidia-management-library-nvml}.} for power measurements. The theoretical results reflect \emph{compute-bound} performance and are thus derived from the theoretical peak computation throughput provided in the datasheets.

\subsection{Digital PIM}

We consider a simple digital PIM architecture consisting of several crossbar arrays that may all operate simultaneously according to the abstract model depicted in Figure~\ref{fig:PIM}(e). We construct the architecture to match the overall GPU memory size of 48GB. The crossbar dimension, per-gate energy, and clock frequency are derived from state-of-the-art digital PIM architectures for memristive~\cite{Nishil, RACER} and DRAM~\cite{SIMDRAM} PIM \revised{that evaluated both the performance characteristics of the underlying devices as well as the necessary periphery and interconnects circuits}, as summarized in Table~\ref{tab:params}. The maximal bitwise throughput supported by such an architecture is the product of the number of rows per crossbar, the number of crossbars in the memory, and the clock frequency. We explore in the code repository the sensitivity to this digital PIM parallelism by varying the crossbar dimensions. The maximum power consumption is derived from the maximal parallelism at full duty cycle. We compare both throughput and normalized throughput per Watt (energy efficiency) \revised{for a power-normalized metric}. The correctness of the algorithms discussed in this paper follows directly from the extensive simulations performed in AritPIM~\cite{AritPIM} and MatPIM~\cite{MatPIM} that both verify correctness (i.e., bit-exact comparison to expected results) and latency (cycles).

\section{High-Throughput Arithmetic}
\vspace{2pt}

We begin by analyzing in this section the elementary benchmark of memory-bound vectored arithmetic. Assume that two $n$-dimensional vectors $\boldsymbol{u}, \boldsymbol{v}$ of $N$-bit numbers (fixed-point or floating-point) reside in the main memory; the goal is to compute an element-wise elementary operation $\circ \in \{+, -, *, /\}$ and store the result as an $n$-dimensional vector $\boldsymbol{z}$ in the memory.

\begin{figure}[!t]
\centering 
\includegraphics[width=\linewidth]{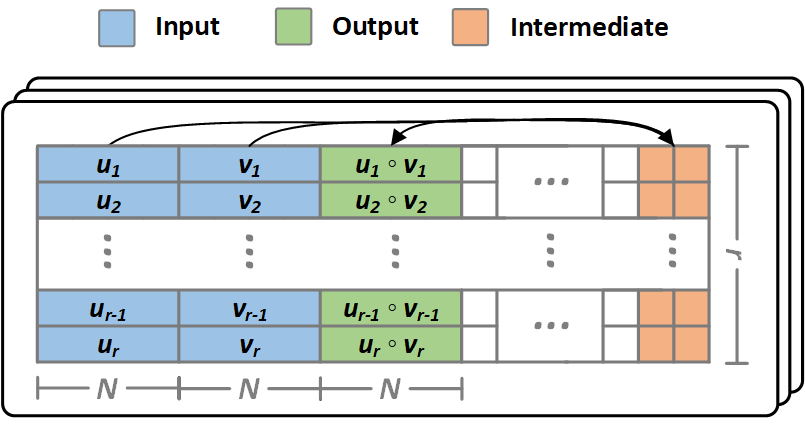}
\caption{The bit-serial element-parallel approach to high-throughput in-memory arithmetic that performs vectored operations in an $r \times c$ crossbar as a sequence of parallel logic operations on columns~\cite{AritPIM}.}
\label{fig:arit} 
\end{figure}

The bit-serial element-parallel approach extends the bitwise parallelism of digital PIM towards maximal arithmetic throughput. Consider an $r \times c$ crossbar with two $r$-dimensional vectors $\boldsymbol{u}, \boldsymbol{v}$ stored with a single $N$-bit element per row (from each vector), as shown in Figure~\ref{fig:arit}. This approach performs the arithmetic in parallel across all rows and all crossbars by constructing the arithmetic operation from a \emph{serial} sequence of logic gates. For example, $N$-bit fixed-point addition is performed by first constructing a $1$-bit full-adder from $9$ serial NOR gates~\cite{AritPIM}, and then performing ripple-carry addition by serially executing $N$ full-adders. While the latency is high at $9 \cdot N = O(N)$ cycles, the throughput is also high at $R/O(N)$ operations per cycle, where $R$ is the total number of rows in the memory (i.e., $r$ multiplied by the number of crossbars). Floating-point operations were originally considered incompatible with digital PIM due to the control flow involved with the alignment and normalization of floating-point numbers. This hurdle was first overcome by FloatPIM~\cite{FloatPIM}; however, several aspects of the design were erroneous~\cite{ReHy, AritPIM}, and it also required the array hardware to support Content Addressable Memory (CAM) functionality. Conversely, AritPIM~\cite{AritPIM} recently proposed a suite of floating-point algorithms that adhere to the IEEE 754 standard without additional hardware. 

\begin{figure}[t]
    \centering
  \begin{tikzpicture}
  
    \begin{groupplot}[
      group style={group size=1 by 2, horizontal sep=1.6cm,
      vertical sep=1.5cm},
      width=7.5cm, height=3.6cm
    ]
    
    \nextgroupplot[
        title={Throughput Comparison},
    	ylabel={\footnotesize Tput. (OPS)},
    	ymode=log,
        enlarge x limits=0.2,
        symbolic x coords={Add, Mult, FP Add, FP Mult},
        xticklabels=\empty,
        xticklabel style={rotate=0},
        ybar, bar width=7pt,
    ]
    \addplot coordinates {(Add,2.33E+14)(Mult,7.41E+12)(FP Add,3.36E+13)(FP Mult,1.16E+13)}; \label{fig:results:arit:MPU}
    \addplot coordinates {(Add,3.49E+11)(Mult,1.11E+10)(FP Add,5.04E+10)(FP Mult,1.74E+10)}; \label{fig:results:arit:DRAM}    
    \addplot coordinates {(Add,5.74E+10)(Mult,5.74E+10)(FP Add,5.74E+10)(FP Mult,5.74E+10)}; \label{fig:results:arit:ExperimentalGPU}   
    \addplot coordinates {(Add,3.87E+13)(Mult,3.87E+13)(FP Add,3.87E+13)(FP Mult,3.87E+13)}; \label{fig:results:arit:TheoreticalGPU}   

    \coordinate (top) at (rel axis cs:0,1);
    
    \nextgroupplot[
        title={Energy Efficiency Comparison},
    	ylabel={\footnotesize Energy Eff. (OPS/W)},
    	ymode=log,
        enlarge x limits=0.2,
        symbolic x coords={Add, Mult, FP Add, FP Mult},
        xtick=data,
        xticklabel style={rotate=25},
        ybar, bar width=7pt,
    ]
    \addplot coordinates {(Add,2.71E+11)(Mult,8.62E+09)(FP Add,3.91E+10)(FP Mult,1.35E+10)};
    \addplot coordinates {(Add,4.43E+09)(Mult,1.41E+08)(FP Add,6.40E+08)(FP Mult,2.21E+08)};
    \addplot coordinates {(Add,1.92E+08)(Mult,1.92E+08)(FP Add,1.92E+08)(FP Mult,1.95E+08)};
    \addplot coordinates {(Add,1.29E+11)(Mult,1.29E+11)(FP Add,1.29E+11)(FP Mult,1.29E+11)};
    
    \coordinate (botl) at (rel axis cs:0,1);

    \coordinate (bot) at (rel axis cs:1,0);
    
    \end{groupplot}
    
    \path (top|-current bounding box.north)--
          coordinate(legendpos)
          (bot|-current bounding box.north);
    \matrix[
        matrix of nodes,
        anchor=south,
        draw,
        inner sep=0.2em,
        draw
      ] at([xshift=-3ex,yshift=2ex]legendpos)
      {
    \ref{fig:results:arit:MPU} & Memristive PIM & [5pt]
    \ref{fig:results:arit:DRAM} & DRAM PIM & [5pt] \\
    \ref{fig:results:arit:ExperimentalGPU} & Experimental GPU & [5pt]
    \ref{fig:results:arit:TheoreticalGPU} & Theoretical Peak GPU & [5pt]
    &[5pt]\\};
    \node[] at ([yshift=4ex,xshift=-8ex]top) {(a)};
    \node[] at ([yshift=5ex,xshift=-8ex]botl) {(b)};
  \end{tikzpicture}
  \caption{Comparison of throughput (operations per second) and normalized throughput per Watt (energy efficiency) for the addition and multiplication of 32-bit fixed-point and 32-bit floating-point (FP) numbers. \revised{For fixed-point addition (multiplication) we find a maximal throughput of 233 TOPS (7.4 TOPS) for memristive PIM, 0.35 TOPS (0.01 TOPS) for DRAM PIM, 0.057 TOPS (0.057 TOPS) for experimental GPU and 38.7 TOPS (38.7 TOPS) for theoretical GPU. Conversely, for floating-point addition (multiplication), we find 33.6 TOPS (11.6 TOPS) for memristive PIM, 0.05 TOPS (0.02 TOPS) for DRAM PIM, 0.057 TOPS (0.057 TOPS) for experimental GPU, and 38.7 TOPS (38.7 TOPS) for theoretical GPU.}}
  \label{fig:results:arit}
\end{figure}

We compare the performance of the suite of arithmetic functions proposed in AritPIM~\cite{AritPIM} to experimental and theoretical GPU performance in Figure~\ref{fig:results:arit}. The experimental GPU performance is bounded by the memory bandwidth for reading the input vectors $\boldsymbol{u}, \boldsymbol{v}$ and writing the result $\boldsymbol{z}$, as indicated by the $>94\%$ DRAM memory bandwidth recorded across all functions. Therefore, it depends only on the bit width of the underlying arithmetic operation. The theoretical compute-bound GPU performance reflects the theoretical throughput provided in an ideal circumstance where memory operations are not required.
For digital PIM, in the spirit of~\cite{Bitlet}, we define the \emph{compute complexity (CC)} as the number of logic gates performed per bit (e.g., $9N/(3 N) = 3$ for $N$-bit fixed-point addition and approximately $10N^2/(4N) = 2.5N$ for $N$-bit fixed-point multiplication with $2N$-bit output), where the number of bits is the total number of input and output bits. Naturally, we find an inverse relationship between CC and the improvement over experimental GPU performance, as shown in Figure~\ref{fig:results:aritCC}. Therefore, PIM is most effective compared to memory-bound GPU when the CC is low. Notice that 16-bit and 32-bit addition possess the same CC as PIM addition latency is linear in $N$, while multiplication CC increases with $N$.

\begin{figure}[t]
    \centering
  \begin{tikzpicture}
    \begin{groupplot}[
      group style={group size=1 by 1, horizontal sep=1.6cm,
      vertical sep=1.3cm},
      width=7.5cm, height=6.2cm
    ]
    \nextgroupplot[
        xlabel={\footnotesize Compute Complexity (CC)},
        ylabel={\footnotesize PIM Tput. / Experimental GPU Tput.},
        every axis plot/.append style={ultra thick},
        every tick label/.append style={font=\footnotesize},
        ymode=log,
        xmin=-10,
        xmax=260
    ]
    \addplot[
        scatter, 
        scatter src=explicit symbolic,
        visualization depends on={value \thisrow{angle} \as \myangle},
        visualization depends on={\thisrow{xshift} \as \myxshift},
        visualization depends on={\thisrow{yshift} \as \myyshift},
        nodes near coords style={
            anchor={\myangle},
            font=\tiny,
            xshift={\myxshift},
            yshift={\myyshift}
        },
        nodes near coords,
        mark=*,
        only marks,
        color=blue!50!white,
    ] table [
        x=x,
        y=y,
        meta=label,
    ] {
        x  y label angle xshift yshift
        6.02E+00 4.07E+03 Add16/32 west 0cm 0cm
        1.03E+02 2.39E+02 Mult16 west -0.05cm 0.175cm
        4.12E+01 5.96E+02 FPAdd16/32 west -0.025cm 0.15cm
        5.79E+01 4.24E+02 FPMult16 west 0.05cm 0.1cm
        1.89E+02 1.29E+02 Mult32 west 0cm 0.125cm
        1.21E+02 2.02E+02 FPMult32 west 0.1cm 0.125cm
    };\label{fig:results:aritCC:MPU};
    \addplot[domain=5:250,color=blue!50!white,samples=200]{4.07E+03/(x/6.02E+00)};
    
    \addplot[
        scatter, 
        scatter src=explicit symbolic,
        visualization depends on={value \thisrow{angle} \as \myangle},
        visualization depends on={\thisrow{xshift} \as \myxshift},
        visualization depends on={\thisrow{yshift} \as \myyshift},
        nodes near coords style={
            anchor={\myangle},
            font=\tiny,
            xshift={\myxshift},
            yshift={\myyshift}
        },
        nodes near coords,
        mark=*,
        only marks,
        color=red!50!white,
    ] table [
        x=x,
        y=y,
        meta=label,
    ] {
        x  y label angle xshift yshift
        6.02E+00 6.10E+00 Add16/32 west 0cm 0cm
        1.03E+02 3.59E-01 Mult16 west -0.05cm 0.175cm
        4.12E+01 8.94E-01 FPAdd16/32 west -0.025cm 0.15cm
        5.79E+01 6.36E-01 FPMult16 west 0.05cm 0.1cm
        1.89E+02 1.94E-01 Mult32 west 0cm 0.125cm
        1.21E+02 3.03E-01 FPMult32 west 0.1cm 0.125cm
    };\label{fig:results:aritCC:DRAM};
    \addplot[domain=5:250,color=red!50!white,samples=200]{6.10E+00/(x/6.02E+00)};
    
    \addplot[
        domain=-10:260, 
        line width=1pt,
        color=gray, 
        dashed, 
        samples=2
    ] {1};
    
    \coordinate (top) at (rel axis cs:0,1);
    
    \end{groupplot}
    
    \path (top|-current bounding box.north)--
          coordinate(legendpos)
          (bot|-current bounding box.north);
    \matrix[
        matrix of nodes,
        anchor=south,
        draw,
        inner sep=0.2em,
        draw
      ] at([xshift=-3ex,yshift=2ex]legendpos)
      {
    \ref{fig:results:aritCC:MPU} & Memristive PIM & [5pt]
    \ref{fig:results:aritCC:DRAM} & DRAM PIM & [5pt] \\
    &[5pt]\\};
    
  \end{tikzpicture}
  \caption{Inverse relationship between compute complexity and improvement over memory-bound GPU (experimental).}
  \label{fig:results:aritCC}
\end{figure}

\section{Matrix Multiplication/Convolution}
\vspace{2pt}

\begin{figure}[t]
    \centering
  \begin{tikzpicture}
  
    \begin{groupplot}[
      group style={group size=1 by 2, horizontal sep=1.6cm,
      vertical sep=1.5cm},
      width=7.5cm, height=3.6cm
    ]
    
    \nextgroupplot[
        title={Throughput Comparison},
    	ylabel={\footnotesize Tput. (OPS)},
    	ymode=log, ymin=1,
        enlarge x limits=0.2,
        symbolic x coords={$32\times 32$, $128\times 128$, $512\times 512$, $2048\times 2048$},
        xticklabels=\empty,
        xticklabel style={rotate=0},
        ybar, bar width=7pt,
    ]
    \addplot coordinates {($32\times 32$,2.63E+08)($128\times 128$,4.11E+06)($512\times 512$,6.42E+04)($2048\times 2048$,1.00E+03)}; \label{fig:results:mat:MPU}
    \addplot coordinates {($32\times 32$,3.94E+05)($128\times 128$,6.16E+03)($512\times 512$,9.63E+01)($2048\times 2048$,1.50E+00)}; \label{fig:results:mat:DRAM}
    \addplot coordinates {($32\times 32$,1.42E+07)($128\times 128$,2.97E+06)($512\times 512$,6.78E+04)($2048\times 2048$,1.03E+03)}; \label{fig:results:mat:ExperimentalGPU}
    \addplot coordinates {($32\times 32$,1.18E+09)($128\times 128$,1.85E+07)($512\times 512$,2.88E+05)($2048\times 2048$,4.51E+03)}; \label{fig:results:mat:TheoreticalGPU}

    \coordinate (top) at (rel axis cs:0,1);
    
    \nextgroupplot[
        title={Energy Efficiency Comparison},
    	ylabel={\footnotesize Energy Eff. (OPS/W)},
    	ymode=log,
        enlarge x limits=0.2,
        symbolic x coords={$32\times 32$, $128\times 128$, $512\times 512$, $2048\times 2048$},
        xtick=data,
        xticklabel style={rotate=25},
        ybar, bar width=7pt,
    ]
    \addplot coordinates {($32\times 32$,3.06E+05)($128\times 128$,4.78E+03)($512\times 512$,7.47E+01)($2048\times 2048$,1.17E+00)};
    \addplot coordinates {($32\times 32$,5.01E+03)($128\times 128$,7.83E+01)($512\times 512$,1.22E+00)($2048\times 2048$,1.91E-02)};
    \addplot coordinates {($32\times 32$,4.93E+04)($128\times 128$,9.89E+03)($512\times 512$,2.27E+02)($2048\times 2048$,3.45E+00)};
    \addplot coordinates {($32\times 32$,3.94E+06)($128\times 128$,6.15E+04)($512\times 512$,9.61E+02)($2048\times 2048$,1.50E+01)};
    
    \draw (axis cs:{[normalized]\pgfkeysvalueof{/pgfplots/xmin}},1)
        -- (axis cs:{[normalized]\pgfkeysvalueof{/pgfplots/xmax}},1);
    
    \coordinate (botl) at (rel axis cs:0,1);

    \coordinate (bot) at (rel axis cs:1,0);
    
    \end{groupplot}
    
    \path (top|-current bounding box.north)--
          coordinate(legendpos)
          (bot|-current bounding box.north);
    \matrix[
        matrix of nodes,
        anchor=south,
        draw,
        inner sep=0.2em,
        draw
      ] at([xshift=-3ex,yshift=2ex]legendpos)
      {
    \ref{fig:results:mat:MPU} & Memristive PIM & [5pt]
    \ref{fig:results:mat:DRAM} & DRAM PIM & [5pt] \\
    \ref{fig:results:mat:ExperimentalGPU} & Experimental GPU & [5pt]
    \ref{fig:results:mat:TheoreticalGPU} & Theoretical Peak GPU & [5pt]
    &[5pt]\\};
    \node[] at ([yshift=4ex,xshift=-8ex]top) {(a)};
    \node[] at ([yshift=5ex,xshift=-8ex]botl) {(b)};
  \end{tikzpicture}
  \caption{Throughput (matrix mult. per second) and throughput per Watt (energy efficiency) comparison for $n \times n$ matrix multiplication with 32-bit floating-point numbers.}
  \label{fig:results:mat}
\end{figure}

This section investigates the performance of matrix multiplication and 2D convolution via digital PIM. The extension of the arithmetic parallelism provided in the previous section to such matrix operations was first investigated in FloatPIM~\cite{FloatPIM} and then generalized in MatPIM~\cite{MatPIM}. These works express the matrix operations as a serial sequence of vectored arithmetic operations, thereby utilizing digital PIM for the vector parallelism. The matrix operations are characterized by high data reuse, and thus the experimental memory-bound GPU approaches compute-bound performance. 

Figure~\ref{fig:results:mat} compares the performance of digital PIM approaches to experimental and theoretical GPU performance for \emph{batched} matrix multiplication on many pairs of matrices (both of dimension $n \times n$). In this operation, we find reuse of $O(n)$ as there is a total of $O(n^3)$ operations operating on $O(n^2)$ data. Therefore, as $n$ increases, we expect the performance gap between experimental memory-bound GPU and theoretical compute-bound GPU will diminish as data movement is no longer the bottleneck. Indeed, we find in Figure~\ref{fig:results:mat} that the gap shrinks with increasing $n$ (e.g., $n=32$ has a significantly larger gap than $n=128$). Overall, for \revised{representation size} $N=32$, we conclude that starting at $n=128$, the experimental GPU energy efficiency surpasses that of digital PIM due to the data reuse mitigating the memory wall bottleneck. Two-dimensional convolution with a $k\times k$ kernel on a $W \times H$ image has similar considerations with data reuse of $O(k^2)$ ($O(WHk^2)$ operations on $O(WH)$ data). 

\section{CNN Inference and Training}
\vspace{2pt}

We culminate the benchmark analysis by evaluating full-precision inference and training of convolutional neural networks (CNN). The acceleration of such neural networks primarily involves matrix multiplication for fully connected layers and 2D convolution for the convolutional layers, as well as element-wise operations for activation functions (e.g., ReLU). We consider a benchmark consisting of the AlexNet, GoogLeNet and ResNet--50 convolutional neural networks with the ImageNet dataset. We evaluate the digital PIM performance by considering only the required matrix multiplication and 2D convolution operations, thereby providing an upper bound on the digital PIM performance that is also close to the true performance \revised{since these operations comprise the majority of the overall compute requirements.} The experimental GPU performance is measured through the PyTorch implementations of the CNNs with random input images of size $224 \times 224 \times 3$. 

\begin{figure}[t]
    \centering
  \begin{tikzpicture}
  
    \begin{groupplot}[
      group style={group size=1 by 2, horizontal sep=1.6cm,
      vertical sep=1.5cm},
      width=7.5cm, height=3.6cm
    ]
    
    \nextgroupplot[
        title={Throughput Comparison},
    	ylabel={\footnotesize Tput. (OPS)},
    	ymode=log,
        enlarge x limits=0.2,
        symbolic x coords={AlexNet, ResNet--50, GoogLeNet},
        xticklabels=\empty,
        xticklabel style={rotate=0},
        ybar, bar width=7pt,
    ]
    \addplot coordinates {(AlexNet,1.19E+04)(ResNet--50,2.00E+03)(GoogLeNet,6.86E+03)}; \label{fig:results:cnn:inference:MPU};
    \addplot coordinates {(AlexNet,1.78E+01)(ResNet--50,3.01E+00)(GoogLeNet,1.03E+01)}; \label{fig:results:cnn:inference:DRAM};
    \addplot coordinates {(AlexNet,1.63E+04)(ResNet--50,1.36E+03)(GoogLeNet,3.49E+03)}; \label{fig:results:cnn:inference:ExperimentalGPU};
    \addplot coordinates {(AlexNet,5.34E+04)(ResNet--50,9.01E+03)(GoogLeNet,3.08E+04)}; \label{fig:results:cnn:inference:TheoreticalGPU};

    \coordinate (top) at (rel axis cs:0,1);
    
    \nextgroupplot[
        title={Energy Efficiency Comparison},
    	ylabel={\footnotesize Energy Eff. (OPS/W)},
    	ymode=log,
        enlarge x limits=0.2,
        symbolic x coords={AlexNet, ResNet--50, GoogLeNet},
        xtick=data,
        xticklabel style={rotate=25},
        ybar, bar width=7pt,
    ]
    \addplot coordinates {(AlexNet,1.38E+01)(ResNet--50,2.33E+00)(GoogLeNet,7.98E+00)};
    \addplot coordinates {(AlexNet,2.26E-01)(ResNet--50,3.82E-02)(GoogLeNet,1.31E-01)};
    \addplot coordinates {(AlexNet,5.55E+01)(ResNet--50,4.43E+00)(GoogLeNet,1.18E+01)};
    \addplot coordinates {(AlexNet,1.78E+02)(ResNet--50,3.00E+01)(GoogLeNet,1.03E+02)};
    
    \draw (axis cs:{[normalized]\pgfkeysvalueof{/pgfplots/xmin}},1)
            -- (axis cs:{[normalized]\pgfkeysvalueof{/pgfplots/xmax}},1);
    
    \coordinate (botl) at (rel axis cs:0,1);

    \coordinate (bot) at (rel axis cs:1,0);
    
    \end{groupplot}
    
    \path (top|-current bounding box.north)--
          coordinate(legendpos)
          (bot|-current bounding box.north);
    \matrix[
        matrix of nodes,
        anchor=south,
        draw,
        inner sep=0.2em,
        draw
      ] at([xshift=-3ex,yshift=2ex]legendpos)
      {
    \ref{fig:results:cnn:inference:MPU} & Memristive PIM & [5pt]
    \ref{fig:results:cnn:inference:DRAM} & DRAM PIM & [5pt] \\
    \ref{fig:results:cnn:inference:ExperimentalGPU} & Experimental GPU & [5pt]
    \ref{fig:results:cnn:inference:TheoreticalGPU} & Theoretical Peak GPU & [5pt]
    &[5pt]\\};
    \node[] at ([yshift=4ex,xshift=-8ex]top) {(a)};
    \node[] at ([yshift=5ex,xshift=-8ex]botl) {(b)};
  \end{tikzpicture}
  \caption{Throughput and normalized throughput per Watt (energy efficiency) for full-precision CNN inference.}
  \label{fig:results:cnn:inference}
\end{figure}

\begin{figure}[t]
    \centering
    \begin{tikzpicture}
    \begin{groupplot}[
      group style={group size=1 by 2, horizontal sep=1.6cm,
      vertical sep=1.5cm},
      width=7.5cm, height=3.6cm
    ]
    
    \nextgroupplot[
        title={Throughput Comparison},
        ylabel={\footnotesize Tput. (OPS)},
        ymode=log,
        enlarge x limits=0.2,
        symbolic x coords={AlexNet, ResNet--50, GoogLeNet},
        xticklabels=\empty,
        xticklabel style={rotate=0},
        ybar, bar width=7pt,
    ]
    \addplot coordinates {(AlexNet,2.58E+03)(ResNet--50,6.72E+02)(GoogLeNet,2.13E+03)}; \label{fig:results:cnn:training:MPU};
    \addplot coordinates {(AlexNet,3.88E+00)(ResNet--50,1.01E+00)(GoogLeNet,3.20E+00)}; \label{fig:results:cnn:training:DRAM};
    \addplot coordinates {(AlexNet,5.78E+03)(ResNet--50,3.96E+02)(GoogLeNet,1.01E+03)}; \label{fig:results:cnn:training:ExperimentalGPU};
    \addplot coordinates {(AlexNet,1.16E+04)(ResNet--50,3.02E+03)(GoogLeNet,9.58E+03)}; \label{fig:results:cnn:training:TheoreticalGPU};
    
    \coordinate (top) at (rel axis cs:0,1);
    
    \nextgroupplot[
        title={Energy Efficiency Comparison},
        ylabel={\footnotesize Energy Eff. (OPS/W)},
        ymode=log,
        enlarge x limits=0.2,
        symbolic x coords={AlexNet, ResNet--50, GoogLeNet},
        xtick=data,
        xticklabel style={rotate=25},
        ybar, bar width=7pt,
    ]
    \addplot coordinates {(AlexNet,3.01E+00)(ResNet--50,7.82E-01)(GoogLeNet,2.48E+00)};
    \addplot coordinates {(AlexNet,4.93E-02)(ResNet--50,1.28E-02)(GoogLeNet,4.06E-02)};
    \addplot coordinates {(AlexNet,1.96E+01)(ResNet--50,1.35E+00)(GoogLeNet,3.45E+00)};
    \addplot coordinates {(AlexNet,3.87E+01)(ResNet--50,1.01E+01)(GoogLeNet,3.19E+01)};
    
    \draw (axis cs:{[normalized]\pgfkeysvalueof{/pgfplots/xmin}},1)
            -- (axis cs:{[normalized]\pgfkeysvalueof{/pgfplots/xmax}},1);
    
    \coordinate (botl) at (rel axis cs:0,1);
    
    \coordinate (bot) at (rel axis cs:1,0);
    
    \end{groupplot}
    
    \path (top|-current bounding box.north)--
          coordinate(legendpos)
          (bot|-current bounding box.north);
    \matrix[
        matrix of nodes,
        anchor=south,
        draw,
        inner sep=0.2em,
        draw
      ] at([xshift=-3ex,yshift=2ex]legendpos)
      {
    \ref{fig:results:cnn:training:MPU} & Memristive PIM & [5pt]
    \ref{fig:results:cnn:training:DRAM} & DRAM PIM & [5pt] \\
    \ref{fig:results:cnn:training:ExperimentalGPU} & Experimental GPU & [5pt]
    \ref{fig:results:cnn:training:TheoreticalGPU} & Theoretical Peak GPU & [5pt]
    &[5pt]\\};
    \node[] at ([yshift=4ex,xshift=-8ex]top) {(a)};
    \node[] at ([yshift=5ex,xshift=-8ex]botl) {(b)};
    \end{tikzpicture}
  \caption{Throughput and normalized throughput per Watt (energy efficiency) for full-precision CNN training.}
  \label{fig:results:cnn:training}
\end{figure}

In Figure~\ref{fig:results:cnn:inference} (Figure~\ref{fig:results:cnn:training}), we compare the upper-bound digital PIM performance with the GPU performance on CNN inference (training) for 32-bit floating-point numbers. We find that the experimental GPU performance is close to the theoretical peak performance across all models due to the moderately high data reuse (e.g., $55-67\%$ L2 hit rate on top of additional reuse at the register level) -- notice that the gap in ResNet and GoogLeNet is more significant than AlexNet since some of their operations are with low reuse (e.g., residual connections, $1\times 1$ convolutions). For the same reason, we find that the digital memristive PIM performance is not significantly better than the GPU performance, and digital memristive PIM energy is slightly worse. 

The code repository includes a sensitivity analysis that evaluates (1) the impact of GPU choice (A100 rather than A6000), (2) quantization with 16-bit floating-point precision, and (3) the impact of varying the PIM parallelism, demonstrating similar trends throughout. 
Overall, we observe the same trend as Figure~\ref{fig:results:cnn:inference} where the gap between the experimental and theoretical GPU is slim, and thus the digital PIM performance is not significantly better than the experimental GPU.

\section{Discussion and Conclusion}
\vspace{2pt}

Recent works have proposed to utilize emerging digital bitwise PIM approaches toward the full-precision acceleration of CNNs as an alternative to analog PIM. This approach benefits from the high accuracy of digital floating point operations and the reduction in data transfer due to the in-memory operations. However, through a series of benchmarks starting with element-wise vectored operations and culminating in CNN inference and training, we demonstrate that digital PIM approaches, with current parameters, are unable to surpass GPU performance for full-precision CNN acceleration. 
These updated results differ from those presented in FloatPIM~\cite{FloatPIM} primarily due to the shift of the model weights from the CPU to the GPU in the baseline evaluation. \revised{Furthermore, additional non-idealities not considered in this work such as spatial/temporal variability and resistance drift only further exacerbate this conclusion.} Therefore, we find that the acceleration of CNN inference and training is not well supported by digital PIM. Rather, it may be better supported by other approaches such as analog PIM architectures as they support significantly higher operation throughput \emph{with reduced accuracy} due to the computation being performed in the analog domain rather than via floating-point arithmetic algorithms.

We identify the poor performance and poor efficiency of digital PIM in full-precision CNN acceleration as arising from a combination of two factors: high compute complexity for the underlying arithmetic operations, and high data reuse in CNN architectures. The analysis in Figure~\ref{fig:results:aritCC} reveals that floating-point multiplication possesses relatively high compute complexity and thus already has a low throughput improvement over GPU performance. Furthermore, Figure~\ref{fig:results:arit} demonstrates that this improvement originates entirely from the memory wall bottleneck throttling the GPU performance as the theoretical compute-bound GPU results surpass digital PIM. 
Therefore, when we increase the data reuse in Figure~\ref{fig:results:mat}, we find that the PIM performance becomes inferior to GPU performance as the memory wall is no longer the bottleneck. That is, we find that it is the combination of the high compute complexity and the high data reuse that leads to the inferior digital PIM performance in full-precision CNN acceleration. \revised{Notice that we believe that this conclusion generalizes to additional PIM architectures as (1) GPU architectures achieve very high compute utilization for operations such as matrix multiplications and convolutions so it is unlikely that any PIM architecture may directly compete with raw GPU compute throughput, and (2) while the exact parameters may differ the overall trends discussed in this paper (of compute complexity and locality) remain agnostic to the specific architecture.}

\begin{figure}[!t]
\centering 
\includegraphics[width=\linewidth, trim={0cm, 0cm, 0cm, 0cm}]{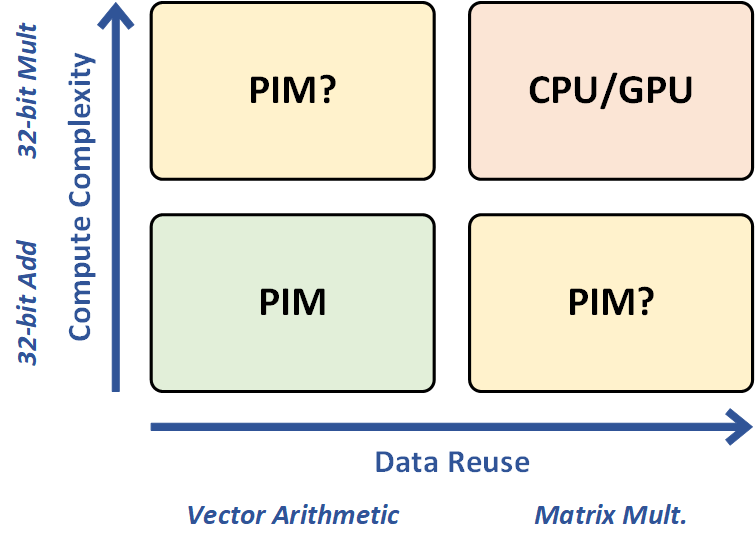}
\caption{Summary of the criteria indicative of PIM performance compared to traditional computing systems.}
\label{fig:summary} 
\end{figure}

\revised{In contrast, applications that possess either low data reuse or low compute complexity have the potential for PIM acceleration. For example, a recent work~\cite{park2024attacc} demonstrated the acceleration of the decoding phase of large-language-models (LLMs) using SRAM-based PIM architectures which were able to accelerate the memory-bound attention algorithm that possesses low data reuse as the keys and value vectors of all previous tokens are compared to a single query. Essentially, this attention algorithm is equivalent to a matrix-vector multiplication which has no data reuse for the matrix.} Hence, as summarized in Figure~\ref{fig:summary}, future work may focus on applications that prioritize arithmetic operations with low compute complexity or low data reuse in GPUs. \revised{Another potential research direction for future works involves the integration of processing-in-memory capabilities in the DRAM memory of the GPU architecture, thereby enabling the usage of PIM algorithms precisely when they are required without requiring additional data-transfer to a separate PIM chip.}

\section{Acknowledgments}
\vspace{2pt}
This work was supported by the European Research Council through the European Union's Horizon 2020 Research and Innovation Programme under Grant 757259, by the European Research Council through the European Union's Horizon Research and Innovation Programme under Grant 101069336, by the Israel Science Foundation under Grant 1514/17, and by the NSF-BSF under Grant 2020-613.

\def\refname{References}
\bibliographystyle{IEEEtran}
\bibliography{refs}

\section*{Biographies}

\paragraph{Orian Leitersdorf}{\,} recently completed his PhD at the Viterbi Faculty of Electrical and Computer Engineering, Technion -- Israel Institute of Technology, Haifa, 3200003, Israel. His current research aims to accelerate digital PIM applications while also addressing challenges such as reliability. Leitersdorf received his M.Sc. degree in Electrical and Computer Engineering from the Technion. 
Contact him at orianl@campus.technion.ac.il.

\paragraph{Ronny Ronen}{\,} is a Senior Researcher at the Viterbi Faculty of Electrical and Computer Engineering, Technion -- Israel Institute of Technology, Haifa, 3200003, Israel. His research interests are in the domain of computer architecture, focusing on processing in memory systems. Ronen received his M.Sc in Computer Science from the Technion. He is a fellow of the IEEE. Contact him at ronny.ronen@technion.ac.il.

\paragraph{Shahar Kvatinsky} {\,} is a Professor at the Viterbi Faculty of Electrical and Computer Engineering, Technion -- Israel Institute of Technology, Haifa, 3200003, Israel. His research interests include emerging non-volatile memory technologies, computer architecture design and circuits, and neuromorphic computing. Kvatinsky received his PhD in Electrical and Computer Engineering from the Technion. He is a fellow member of the IEEE. Contact him at shahar@ee.technion.ac.il.

\end{document}